\documentclass[%
 reprint,
 amsmath,amssymb,
 aps,pra,numerical
]{revtex4-1}

\usepackage{graphicx}
\usepackage{color}
\usepackage{dcolumn}
\usepackage{epstopdf}
\usepackage{bm}
\usepackage[utf8]{inputenc}

\begin{document}

\preprint{APS/123-QED}

\title{Condensation in continuous stochastic mass transport models }
\author{Christos Christou and Andreas Schadschneider}
\affiliation{Institute for Theoretical Physics, University of Cologne, Zülpicher Straße 77, D-50937 Köln, Germany}

\date{\today}

\begin{abstract}
We study the dynamics of condensation for a stochastic continuous mass transport process defined on a one-dimensional lattice. Specifically we introduce three different variations of the truncated random average process. We generalize hereby the regular truncated process by introducing a new parameter $\gamma$ and derive a rich phase diagram in the $\rho-\gamma$ plane where several new phases next to the condensate or fluid phase can be observed. Lastly we use an extreme value approach in order to describe the conditions of a condensation transition in the thermodynamic limit. This leads us to a possible explanation of the broken ergodicity property expected for truncation processes.    
  
\end{abstract}

\maketitle


\section{Introduction}

Condensation is a phenomenon occurring in a plethora of physical systems and through several realizations. The most prominent example is the Bose-Einstein Condensation \cite{Griffin}. In statistical physics the notion of real-space condensation is widely used to describe any process in which a finite fraction of some conserved quantity becomes localized in space. In this context there has been an increasing interest in condensation transitions arising out of equilibrium observed in a variety of works in fields as diverse as traffic models  \cite{Loan,Chowdhury,Harris05}, quantum gravity \cite{Burda}, networks \cite{Krapinsky,Bianconi}, economics \cite{Bouchad,Burda02}, granular materials \cite{Weele}, and mass transport \cite{Waclaw}. It is important to note here that many of these different instances of condensation share common and basic features. The analysis of these properties advanced greatly in the last years through the study of driven diffusive systems and, in particular, of the zero-range process (ZRP) or variants of it \cite{Hanney}.

Our goal in this paper is to expand this study to a substantially diverse manifestation of the stochastic mass transport process. In the usual contexts mentioned above, condensation is a phenomenon observed for systems evolving in a countable state space. Here we are concerned with the condensation transition occurring in continuous stochastic mass transport models such as the random average process (RAP)  \cite{Maj96,Maj00,Krug}. RAP can be grasped as a continuous variant of the zero-range process (ZRP) where instead of particle configurations we are engaged with the distribution of continuous variables, which we will call masses. Processes taking place in a continuous phase space setting have served as a basic model for a variety of physical systems and therefore have been the object of a large number of studies in the last years \cite{Zielen02,Zielen022,Rajesh,Cividini,Cividini2,Kundu,Nossan,Mohanty,Pradhan16,Chakraborti,Pradhan,Levine}. 

We have to note here that they also provide an interesting challenge since in contrast to processes with a countable state space the existence of a unique stationary measure is not given \cite{Spitzer,Sinai,Grosskinsky03} allowing us thus to study systems with broken ergodicity \cite{Zielen023,Luck}. Furthermore the metastability and phase transition properties related to the condensation phenomenon which became recently an object of scientific interest \cite{Grosskinsky,Hirschberg,Landim,Evanswaclaw} are easily implemented in a continuous phase space setting.

Several features arising out of complex dynamics, like the ones encountered in traffic models or server communication, are not sufficiently described by the traditional zero-range process and require a truncation mechanism. Here, as in several papers on this field, we describe the dynamics of the system by introducing a probability density function $\phi (r|m)$ which defines the fraction of the transported mass between different sites. In order to realize this complex behavior a modification of the state-dependent function is necessary. This in turn makes a large set of useful results, arising from a large deviation theory approach, inapplicable. An analytical approach to questions regarding the condensation transition and the nature of the condensate seems thus impossible for truncation models. Fortunately rigorous results related to the condensation transition in the thermodynamic limit can be derived by analyzing the extreme value distribution of the free ARAP model.

The remainder of this paper is organized as follows. In the next section we will briefly introduce the random average process. We will describe hereto the dynamics the free asymmetric random average process and discuss some relevant properties arising from the corresponding factorizable single site mass distribution. This more or less extended recapitulation will be the foundation of the calculations presented in section \textbf{IV}. But before that we will introduce the concept of truncation in section \textbf{III} by presenting and analyzing numerically three different cases of a state-dependent probability density function, $\phi(r|m)$, associated with the truncated asymmetric random average process (TARAP), zero-range random average process (ZRRAP) and the shortened random average process (SRAP).  In section \textbf{IV} we will evaluate the extreme value properties of this distribution and discuss the implications for the thermodynamic limit. The last section is reserved for our final remarks.

\section{Model}

The random average process is a stochastic mass transport system defined on a lattice of $L$ sites, which we fix to be one-dimensional with periodic boundary conditions. This picture is equivalent to that of a particle system defined on a ring as shown in Figure 1. Configurations are denoted by $\textbf{m}(t)=\left\{ m_{k}(t):k\in[0,L], \ \ t\in\mathbb{N}\right\}\in\mathcal{M}_L$ where $m_{k}(t)\in\mathbb{R}$ is the mass at site $k\in\Lambda_L$ at the time point $t$. Each mass can be arbitrarily large and the state space is here given by $\mathcal{M}_L=\mathbb{R}_+^{L}$.

\begin{figure}
\includegraphics[scale=0.4]{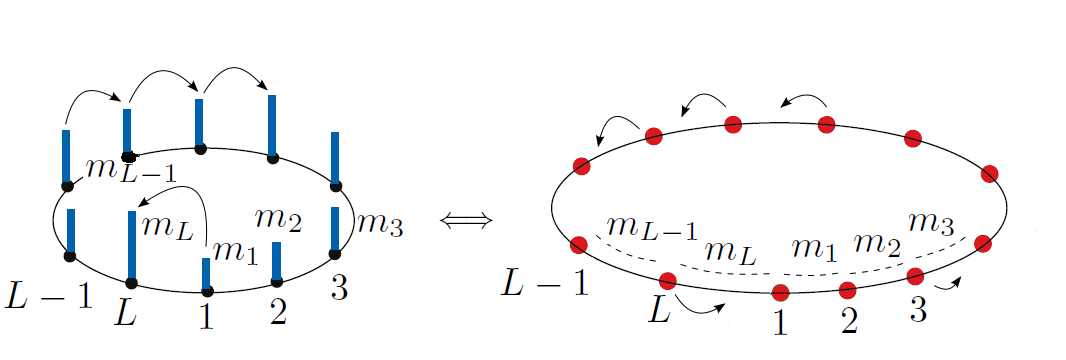}
\caption{Equivalence between the particles on a circle and mass exchange between different sites picture for the asymmetric random average process}
\end{figure}

We consider here parallel update dynamics where at each discrete time step a random vector $\textbf{r}(t)=\left\{ r_{k}(t):k\in[0,L], \ \ t\in\mathbb{N}\right\}$ with $r_{i}(t)\in[0,1]$ is generated from a probability density function $\phi(r|m)=\phi(r_1(t),...|m_1(t)...)$. The random variable $r_{i}(t)$ determines the amount of mass $r_{i}(t)m_{i}(t)$ chipped off from site $i$. For totally asymmetric processes the chipped off mass is transported to one neighboring site (we choose here the direction $i\rightarrow i+1$). The process is therefore described by the equation
\begin{equation}
m_{i}(t)=(1-r_{i}(t-1))m_{i}(t-1)+r_{i-1}(t-1)m_{i-1}(t-1).
\label{MasterEquationFreeARAP}
\end{equation} 
In the past several rigorous results for state-independent functions, $\phi(r)$, have been derived \cite{Nossan,Mohanty,Pradhan16,Chakraborti,Pradhan}. The simplest version of the asymmetric random average process (ARAP) is given for a constant fraction density $\phi(r)= \text{const}$. This process is often referred to as free ARAP since no restriction (except the conservation of the total mass, $M=\sum_{i}m_{i}(t)=\rho L$) is applied.
      
In the following section we prefer to consider the order statistics $\left\{ \ell_{k}(t)\right\}$, where $\ell_{k}(t)$ is the $k$-largest mass at the time-point $t$ in the system, instead of the mass configuration $\left\{m_{i}(t)\right\}$. Hence we use the notation $\ell_{k}(t)$ with
\begin{equation}
\ell_{L}(t)<\ell_{L-1}(t)<...<\ell_{1}(t)=\max\limits_{1\leq i \leq L}m_i(t).
\end{equation}
We will use the largest value, $\ell_1(\infty)$, of the system in the large time limit as an order parameter in order to characterize the different phases of the system. We hereby compare this value to the expectation of the largest value of the free ARAP which we will characterize by $\ell^{*}$. In Section \textbf{IV} we derive an analytical expression for this value. 

As we will see shortly this observable is not sufficient for a full characterization of the phase diagram. Therefore we introduce the flux in the system at a specific time-point $t$ as
\begin{equation}
J(t)=\frac{\sum\limits_{i=1}^L r_i(t)m_i(t)}{M}.
\end{equation}

\section{Truncated processes}

In the past several models have been proposed in order to describe the stochastic transport for many interactive particle systems \cite{Schad}. In many of these models a reduction of the flux is not only explained by the interaction between the different particles but due to the limited transport properties of the system. As example, data transport between different servers is not only limited by the capacities of the servers but also depends on the competency of the connection between them. In order to describe this feature a truncation mechanism proves useful.      

Usually truncation processes as the one presented in \cite{Zielen023} are defined by a cutoff parameter, $\Delta$, according to which the dynamics of the process are specified. A simple example is the \emph{truncated free asymmetric random average process} for which the fraction of transported mass is set to zero if the transported amount is bigger than the cutoff
\begin{equation}
\phi(r|m)=\left[1-R(m)\right]\delta(r)+\Theta\left(R(m)-r\right)
\end{equation} 
where $\Theta$ is the Heaviside step-function and
\begin{equation}
R(m)=\min\left(1,\Delta m^{-1}\right)
\end{equation}
represents the maximum possible fraction.

In the following we choose to set $\Delta=1$. This of course does not lead to a loss of generality since the dynamics of the process and specifically the occurrence of a condensate are fully specified by the ratio $\Delta/ \rho$ and the length of the system. The exact dependence on these parameters will be discussed in full detail in the next section.

Truncated process can be generalized further in order to describe processes for which the expected flux is below that of the free ARAP
\begin{equation}
\langle J \rangle\leq 0.5\rho. 
\end{equation}
This definition deviates from the usual one associated with processes for which the transported fraction of mass is finite even in the thermodynamic limit. We will consider here the special case where the difference $0.5\rho-\langle J \rangle $ depends solely on the actual configuration $m(t)=(m_1(t),...,m_L(t))$ and is expressed through the probability density function of the fractions $\phi(r|m)$. 

We start now with the analysis of truncation processes with a finite length of the lattice. Hereby one has to consider next to the probability of the condensate transition also the stability of those condensates. Unfortunately an analysis of this property proves to be difficult, if not impossible. In order to avoid this problem we decided to introduce a parameter $\gamma$ which allows us to control the persistence of a condensate in the system. The effect of this parameter can be read from the corresponding probability density function $\phi(r|m)$ presented in the following three subsections. 

\subsection{TARAP} 

We start by generalising the case of the truncated random average process (TARAP) by introducing the fraction density    
\begin{equation}
\phi (r|m)=[1-R(m)]\delta(r)+\Theta(R(m)-r),
\label{phiTARAP}
\end{equation} 
where $\Theta$ is the Heaviside step-function and
\begin{equation}
R(m)=\min\left\{1,m^{-\gamma}\right\}
\label{R(m)}
\end{equation}
represents the maximum possible fraction. If we choose to set $\gamma=1$ then the expression of the original TARAP for $\Delta=1$ is restored. The moments of the transported mass per site and time-step, $\mu_k(m)=\langle r^km^k \rangle$, are given by the equation
\begin{equation} 
\mu_k=\int\limits_{0}^{1} \, \mathrm{d}r \, r^km^{k}\phi(r|m)=\frac{m^kR^{k+1}(m)}{k+1}.
\label{meanTARAP} 
\end{equation} 

\begin{figure}
\includegraphics[scale=0.4]{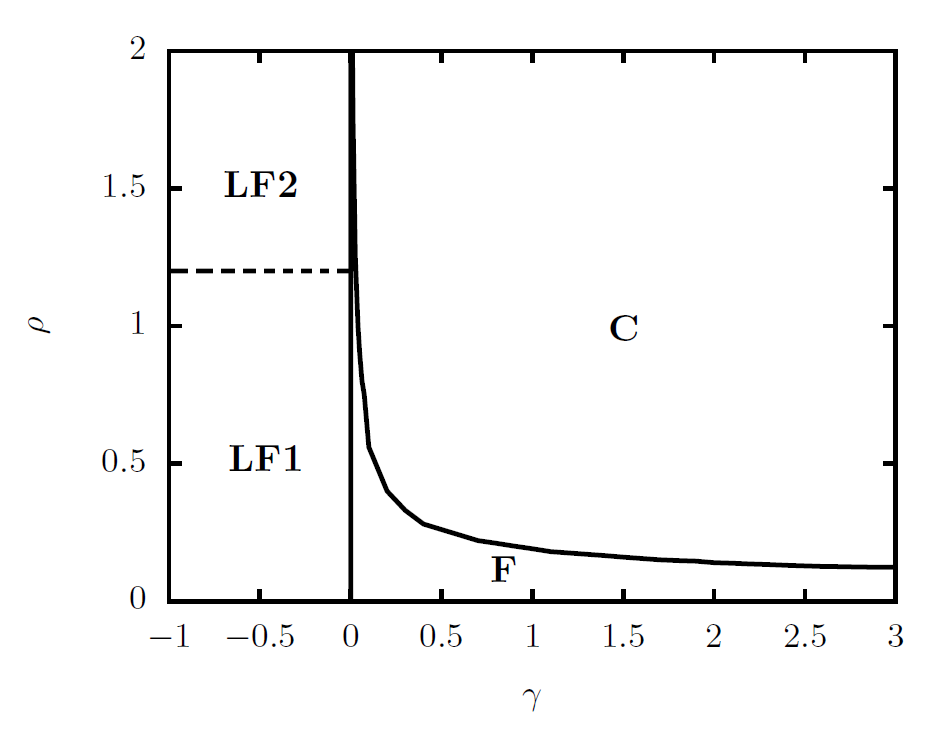}
\caption{Phase diagram for the TARAP model in the $\rho$-$\gamma$ plane for $L=100$. The dashed line does not correspond to a phase transition but describes the crossover between \textbf{LF1}-\textbf{LF2}.}
\label{phasesofTARAP}
\end{figure}

 \begin{figure}
	\input{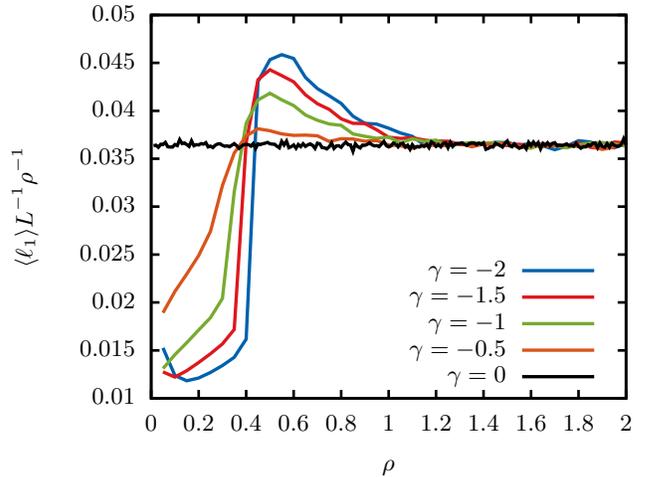}
	\caption{Mean largest value of the TARAP in the stationary limit for $L=100$ and different values of $\gamma$.}  
	\label{TARAPgminus}
\end{figure}

 \begin{figure}
	\input{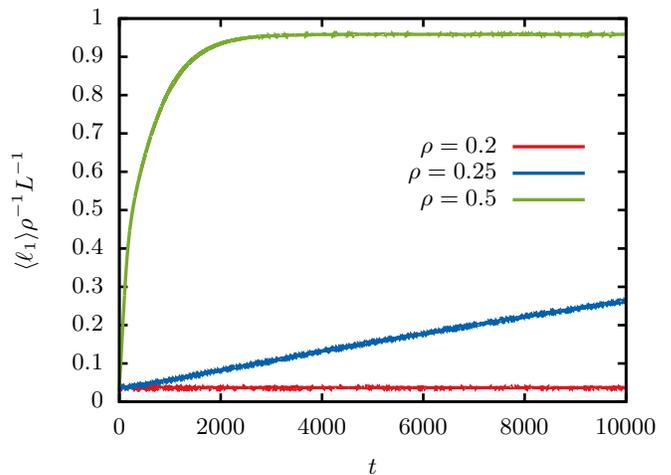}
	\caption{Time evolution of the largest mass for the TARAP for $L=100$, $\gamma=1$ and different densities.}
		\label{TARAPtime}
\end{figure} 

By using Monte Carlo simulations one can observe three different phases as represented in Figure \ref{phasesofTARAP}. We characterize these phases by the distribution of the parameters $\langle \ell_{1}(T)\rangle\rho^{-1}L^{-1}$ and $\langle J(T) \rangle$ averaged over $10^5$ realizations in the stationary regime. We have to note here that since it is impossible to define analytically the time necessary for the distribution to become stationary, a numerical method becomes indispensable. For most of the parameter sets $\left\{\rho,\gamma\right\}$ and $L=100$ a running time of $T=10^5$ is sufficient.

In the following we will consider only systems with $L=100$. This makes the comparison between the different models easier. For $L=100$ we see with a simple Monte-Carlo simulation that the mean largest value for the free ARAP is equal $\ell^{*}=0.0365$. This value will serve as reference point for the characterization of the phases in the following. 

For $\gamma<0$ we see that the flux of the system is below the free ARAP expectations and the overall variance of the single site mass distribution is also minimal. We therefore characterize this phase as low-flow phase (\textbf{LF}). In this phase for small densities the mean largest value of the system is below $\ell^{*}$. This feature changes completely for higher densities ($\rho\approx 0.4$ for $\gamma=-1$), where we observe values of $\ell_1$ well above the free ARAP expectation $\ell^{*}$. We see that in this case the truncation has a significant effect on the distribution of the largest value (\textbf{LF1}).

For increasing densities ($\rho>1.2$) the order parameter $\ell_1$ shows no difference to the free ARAP case as shown in Figure \ref{TARAPgminus}. For such high densities more than half of the sites have a mass above $1$ and the effect of the truncation in the distribution of the largest value becomes negligible. At the same time this state is completely distinguishable  from the free ARAP state since the flux in the system is well below $\rho /2$ (\textbf{LF2}). 

For $\gamma>0$ we can observe two different phases. For small values of $\gamma$ the differences to the free ARAP system are minimal. This phase is often characterized, due to the high flow in the system, as fluid phase (\textbf{F}). The probability of condensation as well as the lifetime of the corresponding condensates are small in this phase and consequently no deviation of $\langle \ell_1 \rangle$ from $\ell^{*}$ could be observed in our simulation.

For high densities and $\gamma$ values  we are in the condensate phase (\textbf{C}). In this phase several condensates may appear in the system and a steady increase of the largest value can be observed for high values of $\rho$ (Figure \ref{TARAPtime}). Different condensates compete with each other due to the conservation of the density and with evolution of time only one condensate may survive.

Regarding the stability of the condensate we can say that the fluctuations observed early in the system diminish for $\ell_1>0.9\rho L$ but never disappear and in rare cases may lead to the destruction of the condensate which is characterized by a drop of the largest mass in the system and the following change in the position of the condensate. Finally we have to note that all of these dynamics observed in the condensate phase are accelerated for lower densities.   

We summarize the characteristics of these phases in the next table.

\begin{table}[h]
\begin{tabular}{ l | c | r }
    & $\langle \ell_1(\infty) \rangle$ & $\langle J(\infty) \rangle$  \\
  \hline			
  \textbf{C} & $> \ell^{*}$ & $\sim 0$  \\
  \hline
  \textbf{F} & $\sim \ell^{*}$ & $\sim \rho /2$ \\
  \hline  
  \textbf{LF1} & strong fluctuations & $< \rho /2$  \\
  \hline 
  \textbf{LF2} & $\sim \ell^{*}$ & $< \rho /2$  \\
  \hline 
\end{tabular}
  \caption{Characterization of the different phases of the TARAP.}
\end{table}

\subsection{ZRRAP}

The truncation of the TARAP analyzed in the last subsection manifests itself in two different ways: the prohibition of transport expressed through the term $\left( 1-R(m)\right)\delta (r)$ and the reduction of the transported fraction as dictated by the term $\Theta \left( R(m)-r\right)$. In order to understand the relevance of these two distinct effects for the condensation mechanics we decided to study two different processes designed according to these terms.

\begin{figure}
\input{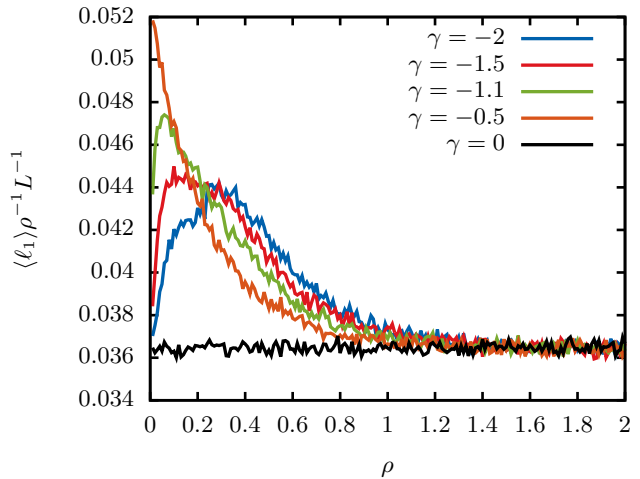}
\caption{Mean largest value of the ZRRAP in the stationary limit for different values of $\gamma$.}
\label{ZRRAP}
\end{figure}

By starting with the first term we arrive at a process which can be regarded as the continuum state space analogue of the zero range process. We will therefore here call this process zero-range random average process (ZRRAP). 
The fraction density of this process is given by 
\begin{eqnarray}
\phi(r|m)=\left(1-R(m) \right)\delta(r)+R(m)\Theta(1-r)
\label{ZRRAPpdfrate}
\end{eqnarray}
where we use for $R(m)$ the same expression as defined above by Eq. (\ref{R(m)}). With regard to the moments of this density we get the expression
\begin{equation} 
\mu_k=\int\limits_{0}^{1} \, \mathrm{d}r \, r^km^{k}\phi(r|m)=\frac{m^k}{k+1}R(m).
\label{meanZRRAP} 
\end{equation}

This fraction density describes a process for which with a state-dependent probability the transport rate $r_{i}(t)$ of a certain site $i$ at the time point $t$ will be set to $0$, while in the complementary event, the rate $r_{i}(t)$ is randomly distributed with equal probability in the interval $[0,1]$. This property is the big difference between the ZRRAP and the TARAP, introduced in the last subsection, where the transported mass per site and time-step had an upper bound of $mR(m)$, as seen by Eq. (\ref{phiTARAP}).

For the ZRRAP three different phases can be defined. By choosing $\gamma<0$ we can see that the mean largest value shows a very interesting behavior, whereas the expected value is always bigger than $\ell^{*}$ for $\rho<1.2$ (\textbf{LF1}). This fact can be observed in Figure \ref{ZRRAP}. Initially a spatial concentration of masses with $m_i>1$ takes place, which travels through the system unhindered. At the same time the rest of the mass in the system contributes only marginally to the overall flow leading thus to a low-flow phase. This local concentration of mass leads of course to a stationary mean largest value with $\langle \ell_1(\infty) \rangle>\ell^{*}$. For higher densities ($\rho>1.2$) this effect disappears and the free ARAP-like case is restored (\textbf{LF2}). 

\begin{figure}
\includegraphics[scale=0.4]{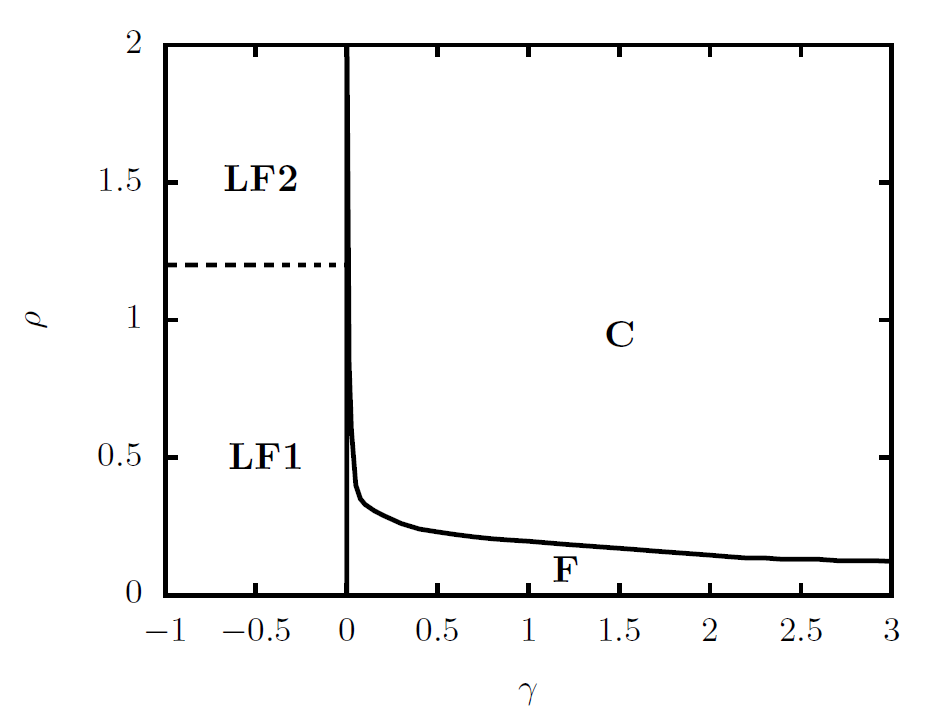}
\caption{Phase diagram for the ZRRAP model in the $\rho$-$\gamma$ plane for $L=100$. The dashed line does not correspond to a phase transition but describes the crossover between \textbf{LF1}-\textbf{LF2}.}
\label{phaseofZRRAP}
\end{figure} 

By studying the system for $\gamma>0$ we can observe again a fluid and condensate phase as for the TARAP (see Figure \ref{phaseofZRRAP}). There is in general a striking similarity between the two diagrams, which proves the relevance of the restraint on the transition ($\delta$-term in Eq. (\ref{phiTARAP})) for the condensation effect. The dynamics of condensation are in general simillar to the ones for the TARAP. The main difference lies in the speed with which the condensate builds up in the system. The concurrence between different condensates resolves hereby faster in the ZRRAP in comparisson to the TARAP. This also reflected in the fact that for the same values of $\gamma$ a smaller $\rho$ is necessary for a condesnate to appear in the system. Fluctuations persist hereby for all parameter values and decrease for increasing values of $\gamma$ and decreasing densities allowing therefore for a smooth crossover to the stable high flow phase (\textbf{F}). 

All of these results lead us to the following tabular representation of the phase diagram. 

\begin{table}[h]
\begin{tabular}{ l | c | r }
    & $\langle \ell_1(\infty) \rangle$ & $\langle J(\infty) \rangle$  \\
  \hline			
  \textbf{C} & $> \ell^{*}$ & $\sim 0$  \\
  \hline
  \textbf{F} & $\sim \ell^{*}$ & $\sim \rho /2$ \\
  \hline  
  \textbf{LF1} & $> \ell^{*}$ & $< \rho /2$  \\
  \hline 
  \textbf{LF2} & $\sim \ell^{*}$ & $< \rho /2$  \\
  \hline 
\end{tabular}
  \caption{Characterization of the different phases of the ZRRAP.}
\end{table}

\subsection{SRAP}

In the last two subsections we investigated the effect of truncation for fraction densities with a non-zero probability for the event $\{ r_{i}(t)=0 \}$. In this subsection we will show that even for transport processes for which $\left\{r_{i}(t)>0 \ \ \forall i,t\right\}$ holds, a condensate phase appears in the $\rho-\gamma$ plane.
 
We introduce hereto the function
\begin{equation}
\phi(r|m)=\frac{1}{R(m)}\Theta\left(R(m)-r\right) 
\label{pdfrate}
\end{equation}
where $\Theta(x_0-x)$ is the Heaviside function and $R(m)$ as in Eq. (\ref{R(m)}) . It is clear that for $\gamma=0$ the free ARAP model is restored. As before the introduction of the parameter $\gamma$ allows us to study systems with a large range of moments
\begin{equation}
\mu_k=\int\limits_{0}^{1} \, \mathrm{d}r \, r^km^{k}\phi(r|m)=\frac{m^kR^k(m)}{k+1}
\label{meanTRAP}
\end{equation}

\begin{figure}
	\includegraphics[scale=0.3]{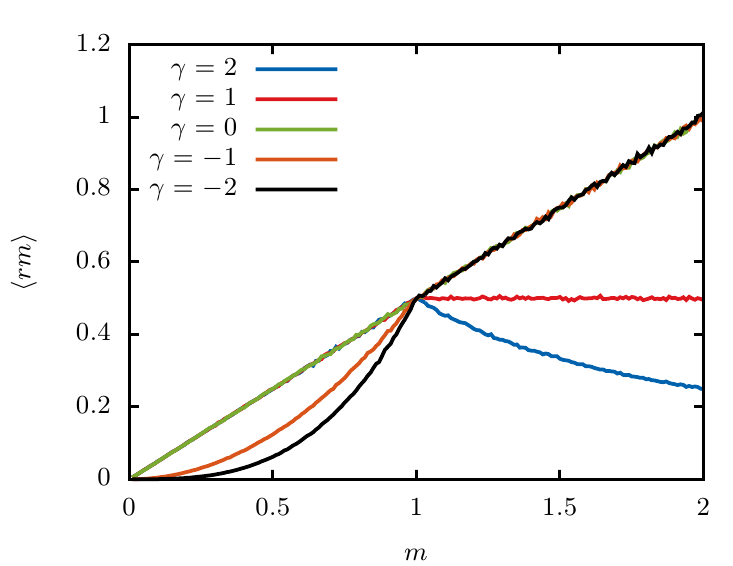}
	\caption{Mean transported mass per time-step for a site with mass $m$ for the ZRRAP and the SRAP and different parameters $\gamma$. We can see that for these two models the first moment $\mu_1$ is described by the same function (Eq. (\ref{meanZRRAP}) and (\ref{meanTRAP})).} 
	\label{model}
\end{figure}

\begin{figure}
\input{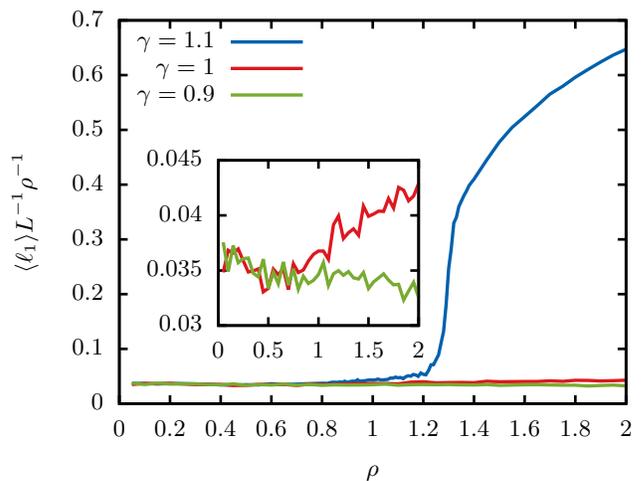}
\caption{Mean largest value  of the SRAP in the stationary limit for different values of $\gamma$.}
\label{SRAP}
\end{figure}

In Figure \ref{model} we can see the first moment of the transported mass $\mu_1$ as function of the mass. We will see later that although the first moments for the transported mass of the two models (SRAP and ZRRAP) are equal, the corresponding phase diagrams have a completely different structure.

The phase diagram of this model differs greatly from those of the other two models. We can see that the condensate/fluid separation appears in the system only for $\gamma>1$. For the fluid case we again observe that the largest value in the system remains constant around the expected value of $\ell^{*}$ with small fluctuations. The same holds for the flux which fluctuates around $0.5\rho$.

In the condensate phase this no longer holds. We can clearly see that $\langle \ell_1(\infty) \rangle> \ell^{*}$ and $\langle J(\infty) \rangle <0.5\rho$. It is especially interesting that in this regime the spatially extended condensate performs a drift through the system. For high values of $\gamma$ $(\gamma>1.5)$ the position of the condensate stabilizes and the largest value in the system rises up to a stationary value which is close to $M$. Fluctuations are present even after the formation of the condensate, but this value remains stationary when integrated over long time-intervals $\left( T=10^3\right)$.

For $0<\gamma<1$ the flow of the system is below $\rho /2$ while at the same time we can observe a homogeneous mass distribution. We therefore characterize this phase as homogeneous low flow phase (\textbf{HLF}). Surprisingly we found that for $\rho>0.9$ as shown in Figure \ref{SRAP} the mean largest value in the system decreases with increasing densities (\textbf{HLF2}). This discovery is explained by the slowdown of the mass drift for sites with $m>1$ due to the truncation effect. Correspondingly we get an equilibration of the mass distribution and thus a lower mean largest value. This effect does not arise for lower densities and we regain the expected mean largest value of $\ell^{*}$ (\textbf{HLF1}).

\begin{figure}
\includegraphics[scale=0.4]{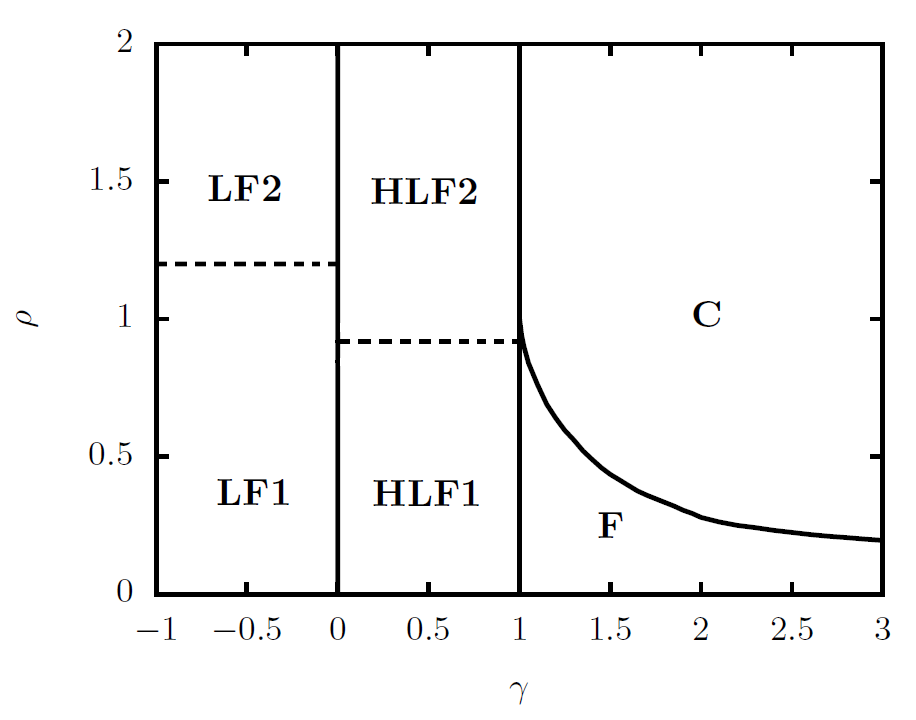}
\caption{Phase diagram for the SRAP model in the $\rho$-$\gamma$ plane for $L=100$. The dashed lines do not correspond to a phase transition but describe the crossover between \textbf{LF1}-\textbf{LF2} and \textbf{HLF1}-\textbf{HLF2} correspondingly.}
\end{figure} 

The phase diagram for this process is quite more complicated for the previous models. But as seen in the table above a clear distinction between the different phases can be made. We have to note here that although for the \textbf{HLF1} and the \textbf{LF2} the truncation effect seems to have no effect on the mean largest value distribution the two phases have completely different characteristics. In the one case $\left(\textbf{LF2}\right)$ we have collective dynamics while in the other $\left(\textbf{HLF1}\right)$ the mean largest value distribution is solely defined by single site mass deviations. This is the reason why these two phases appear on completely different ranges of the density parameter. 

Regarding the $\gamma<0$ case we have the same structure as described above for the TARAP. For $\rho<1.2$ we see again a strong fluctuation of the mean largest value with respect to the density. Like in the former processes these fluctuations disappear for $\rho>1.2$. It becomes clear that the dynamics of condensation for the TARAP and therefore the transition from fluid to the condensate phase are controlled by single site fluctuations and consequently by the prohibition of transport $\left(\delta-\text{term}\right)$. At the same time the nature of the phases for $\gamma<0$ is controlled by collective dynamics and the reduction of the mass transport $\left( \Theta-\text{term}\right)$. 

\begin{table}
\begin{tabular}{ l | c | r }
    & $\langle \ell_1(\infty) \rangle$ & $\langle J(\infty) \rangle$  \\
  \hline			
  \textbf{C} & $> \ell^{*}$ & $\sim 0$  \\
  \hline
  \textbf{F} & $\sim \ell^{*}$ & $\sim \rho /2$ \\
  \hline  
  \textbf{LF1} & strong fluctuations & $< \rho /2$  \\
  \hline 
  \textbf{LF2} & $\sim \ell^{*}$ & $< \rho /2$  \\
  \hline 
  \textbf{HLF1} & $\sim \ell^{*}$ & $< \rho /2$  \\
  \hline 
  \textbf{HLF2} & $\leq \ell^{*}$ & $< \rho /2$  \\
  \hline 
\end{tabular}
  \caption{Characterization of the different phases of the SRAP. No difference between the \textbf{HLF2} and \textbf{LF2} phase can be determined if we observe only the parameters $\ell(\infty)$ and $J(\infty)$. But the two phases have completely different dynamics as explained in the text below.}
\end{table}

\section{Thermodynamic Limit}

In the last section we derived the phase diagram for three different truncation models with finite lengths. One question that arose was if this phase structure survives in the thermodynamic limit. Here we will concentrate on this question by focusing on the transition between the condensate and the fluid phase for increasing lengths.   

For all systems considered in this paper we follow the free ARAP update rules as long as $\ell_{1}(t)<1$ for $\gamma>0$ or $\ell_{L}(t)>1$ correspondingly if $\gamma<0$. This allows us to make some analytical predictions about the behavior of the system in the thermodynamic limit by using some well known results of the past. In the following we use for the values of the single site mass and the extreme value distribution of the free ARAP the notation $m_i$ and $\ell_k$ correspondingly.

As shown in \cite{Evans04} the steady state distribution of mass transport processes factorizes if the fraction density is given by a relation of the form
\begin{equation}
\phi(r|m)\propto v(r)u(m-rm).
\label{factorazible}
\end{equation} 
For this fraction density we have a stationary state described by the mass distribution
\begin{equation}
P(m_1,...,m_L)=\frac{\prod_{i=1}^{L}f(m_i)}{Z(M,L)}\delta\left(\sum\limits_{i=1}^{L}m_i-M\right)
\end{equation}
where the function $f(m)$ is given by
\begin{equation}
f(m)= \int_{0}^{1} \, dr \, mv(r)u(m-rm)
\end{equation} 
and the "canonical partition function" is just the normalization
\begin{equation}
Z(M,L)=\prod\limits_{i=1}^{L}\int_{0}^{\infty} \, \mathrm{d}m_i \, f(m_i)\delta\left(\sum\limits_{i=1}^{L}m_i-M\right)
\end{equation}

We note here that $P(m_1,...,m_L)$ is equivalent to the probability density of picking $L$ independent and identically distributed random variables from a common distribution $f(m)$, conditioned on the fixed value of their total sum. This allows us to verify our analytical results by using a Monte Carlo simulation where we create $L$ independent and identically distributed random variables, $\tilde{m_i}\ \ \forall i \in\mathbb{Z}\cap[0,L]$, with the single-site mass distribution 
\begin{equation}
p(\tilde{m})=f(\tilde{m})\frac{Z\left(\rho L-\tilde{m},L-1\right)}{Z\left(\rho L,L\right)} 
\end{equation}
and normalize them by their sum
\begin{equation}
m_i=\frac{\rho L \tilde{m_i}}{\sum_{i=1}^L \tilde{m_i}}.
\end{equation}
By comparing this method with the more time consuming one, where different initial configurations were evolved according to the update rules derived from the equation (\ref{MasterEquationFreeARAP}), we can observe an excellent agreement. The presented approach is a reasonable approximation in the fluid phase and serves as a simple algorithm to derive the numerical results presented in this section. For processes with truncation such a method would be misleading.

For general fraction densities (\ref{factorazible}) is not fulfilled and a factorization of the corresponding partition function is impossible. Consequently a relation for the critical density, $\rho_c$, of the form
\begin{equation}
\rho_c=\int \, \mathrm{d}m \, f(m)m
\end{equation}
like in \cite{Evans05} cannot be applied. We characterize here as critical density $\rho_c$ the density value above which condensation occurs in the system. 

We know from \cite{Evans06} that the probability of $\ell_{1}\leq x$ is given by 
\begin{equation}
P(x,M,L)=\frac{I(x,M,L)}{Z(M,L)}
\label{probability}
\end{equation}
where
\begin{equation}
I(x,M,L)=\prod\limits_{i=1}^{L}\int_{0}^{x} \, \mathrm{d}m_i \, f(m_i)\delta\left(\sum\limits_{i=1}^{L}m_i-M\right)
\end{equation}
and $Z(M,L)$ is given by $Z(M,L)=I(\infty ,M,L)$ (see equation (5) above).

The Laplace transform of $I(x,M,L)$ is easily computed by
\begin{equation}
\int_{0}^{\infty} \, \mathrm{d}M \, I(x,M,L)e^{-sM}=\left[\int_{0}^{x} \, \mathrm{d}m \, f(m)e^{-sm}\right]^L.
\label{LaplatransformI(x,M,L)}
\end{equation}

\begin{figure}
	\input{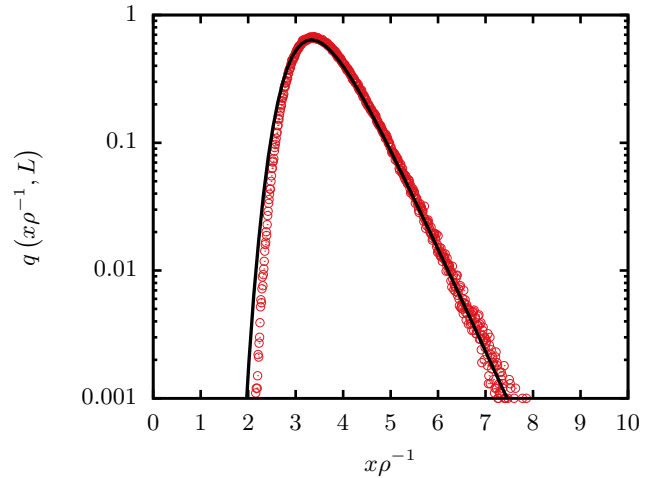}
	\caption{Numerical (red circles) and analytical (black line) derivation of the largest value distribution density for a free ARAP system with $L=100$ and $\rho=1$. } 
	\label{fig1}
\end{figure}

The critical density for a condensation phase transition of the free ARAP is infinite and we can apply the Bromwich integral in order to invert the expression (\ref{LaplatransformI(x,M,L)}) in order to get
\begin{equation}
P(x,M,L)=\frac{\int_{c-i\infty}^{c+i\infty} \, \mathrm{d}s \, \exp \left[L\left(\rho s+\ln g(s)\right)\right]}{\int_{c-i\infty}^{c+i\infty} \, \mathrm{d}s \, \exp\left[ L\left(\rho s-2\ln s\right)\right]}
\end{equation}
with $M=\rho L$ and 
\begin{equation}
g(s)=s^{-2}(1-e^{-sx}-sxe^{-sx}).
\end{equation}
The integration is performed along the vertical line $\Re(s) = c$ in the complex plane such that $c$ is greater than the real part of all singularities of the integrand. Since we are in the fluid phase we can use a saddle point approximation as the one presented in \cite{Majextr} where to leading order the saddle point, $s_0$, of the integrand is independent of $x$ and is given by the equation
\begin{equation}
\rho=\frac{\int_0^{\infty} \, \mathrm{d}m \, m^2e^{-s_0m}}{\int_0^{\infty} \, \mathrm{d}m \, me^{-s_0m}}=\frac{2}{s_0}.
\end{equation}
Inserting this formula in 
\begin{equation}
P(x,M,L)=\exp \left[-L\frac{\int_{x}^{\infty} \, \mathrm{d}m \, me^{-s_0m}}{\int_{0}^{\infty} \, \mathrm{d}m \, me^{-s_0m}} \right]
\end{equation}
leads finally to the approximative solution
\begin{equation}
P\left(x,M,L\right)=P\left(x\rho^{-1},L\right)=\exp \left[-L\left(\frac{2x}{\rho}+1\right)e^{-\frac{2x}{\rho}}\right].
\end{equation}

Although the derived equations are approximations they can be still useful even for finite systems as shown in Figure \ref{fig1} where we can see a good agreement between the Monte Carlo simulation and our analytical prediction for the probability density function of the largest value 
\begin{equation}
q\left(x\rho^{-1},L\right)=\rho\frac{\partial}{\partial x}P\left(x\rho^{-1},L\right).
\end{equation} 
We can use now this expression to calculate the mean largest value in the system 
\begin{eqnarray}
\frac{\langle \ell_{1} \rangle}{\rho}&=\int\limits_0^{\infty} \, \mathrm{d}y \, y q\left(y,L\right)=\rho^{-1}\int\limits_{0}^{\infty} \, \mathrm{d}x \, x \frac{\partial}{\partial x}P\left(x\rho^{-1},L\right)=\nonumber\\
&=-\sum\limits_{n=1}^{\infty}\left(\frac{-L}{n}\right)^n\frac{1}{2n}\sum\limits_{k=0}^{n}\frac{n^k}{k!}.
\label{numintegral}
\end{eqnarray}

\begin{figure}
	\input{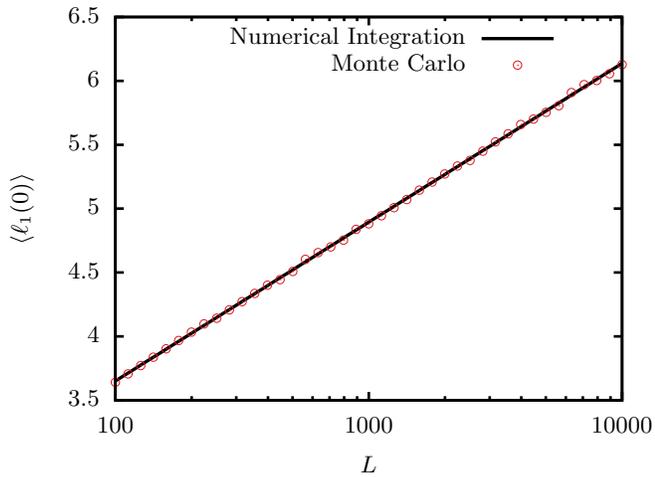}
	\caption{Mean largest value for different lengths in the special case of $\rho =1$. Each data point was calculated by averaging over $10^4$ different Monte Carlo simulations. We have evolved hereby a periodic boundary system with random initial condition according of the dynamics of the free ARAP. The blue line was derived by a numerical evaluation of the integral in Eq. (\ref{numintegral}). }  
	\label{logL}
\end{figure}

We preferred here a numerical evaluation of the integral in Eq. (\ref{numintegral}). The results of which are presented in Figure \ref{logL} where we also plotted the results of a Monte Carlo of a free ARAP system. The derived curve is best approximated by the function 
\begin{equation}
\langle \ell_{1} \rangle=0.54\rho\ln (8.63L+1).
\label{meanlargestvalueformula}
\end{equation} 
Now if we set $L=100$ then we can derive 
\begin{equation}
\langle \ell_1 \rangle=3.65\rho=100\rho\ell^{*} , 
\end{equation}
whereas $\ell^{*}$ is the value for $\langle \ell_1 \rangle \rho^{-1}L^{-1}$ that we could observe in the fluid phase for all of the truncated models studied in the last section. Unfortunately the extreme value distribution can be determined analytically only for the fluid case. 

Nevertheless the presented approximations prove extremely useful in the analysis of the broken ergodicity property of truncated models. This characteristic was first noticed in \cite{Zielen023} where the divergence of the lifetimes of the high flow and the low flow states in the thermodynamic limit could be observed. Here we also expect an ergodicity breaking for the TARAP and the ZRRAP in the case of $\gamma>0$ and for the SRAP if we assume that $\gamma>1$.

This statement relies on two facts, the stability of condensates for truncated models and the degeneracy of the exteme value distribution in the thermodynamic limit. 

In the last section we characterized the condensate phase by the large deviation of the mean largest value from the derived expectation of Eq. (\ref{meanlargestvalueformula}). By using Monte Carlo simulations we could see that for increasing times this deviation is also increasing. Specifically we could say that for large time intervals following property becomes evident
\begin{equation}
\left\langle \frac{\partial \ell_{1}(t)}{\partial t} \right\rangle \geq 0\ \ \ \text{as long as }\ \ \ \ell_{1}(t)>1. 
\end{equation}
This assumption would of course not hold for all time-points in a finite system due to the conservation of the mass, $M$, but it becomes reasonable if we consider the case of $L\rightarrow\infty$. Now for $\ell_1(t)\gg 1$ this would lead to a condensate with infinite lifetime, meaning that the probability of the mass on this site to return to values below $1$ is becoming zero. This condition alone is not sufficient for the appearance of condensates in the studied systems since no guarantee for the survival of states with $\ell_1(t)=1+\varepsilon$ can be made. Therefore we have to consider the order statistics of the free ARAP.  

We note here that the position of the largest value in the fluid phase is neither stable nor does it perform a continuous drift but shows irregular jumps. It is therefore appropriate to consider this as a resetting of the largest value to a random position constantly during the evolution of the system. Due to this resetting it is important to calculate the distribution of the second largest value in order to describe the properties of the transition $\{ \ell_{1}(t)<1\}\leftrightarrow\left\{\ell_{1}(t+1)>1\right\}$. 

We start therefore by the formula
\begin{equation}
\text{Pr}\left\{\ell_{2}<x\right\}=\text{Pr}\left\{\ell_{1}<x\right\}
+\text{Pr}\left\{\ell_{2}<x<\ell_{1}\right\}.
\end{equation} 
Since the term $\text{P}\left\{\ell_{1}<x\right\}$ has been calculated above we concentrate now on the second term \cite{Order}
\begin{eqnarray}
&\text{Pr}\left\{\ell_{2}<x<\ell_{1}\right\}=L\int_{x}^{\infty} \, \mathrm{d}m_1 \, f(m_1)\times\nonumber\\
 &\frac{\prod\limits_{i=2}^{L} \, \int_{0}^{x} \, \mathrm{d}m_i \, f(m_i)\delta\left(\sum\limits_{i=2}^{L}m_i-M-m_1\right)}{\prod\limits_{i=1}^{L}\int_{0}^{\infty} \, \mathrm{d}m_i \, f(m_i)\delta\left(\sum\limits_{i=1}^{L}m_i-M\right)}.
\label{Pl1xl2}
\end{eqnarray}  
In order to evaluate this expression we use, as before, a saddle point approximation of the inverse Laplace transform by determining the minimum of the function
\begin{eqnarray}
h(s)=\rho s&+\frac{1}{L}\ln \int_{x}^{\infty} \, \mathrm{d}m \, f(m)e^{-sm}+\nonumber\\
&+\frac{L-1}{L}\ln \int_{0}^{x} \, \mathrm{d}m \, f(m)e^{-sm}.
\label{saddlepointsecondlargest}
\end{eqnarray} 
Using a numerical calculation we can see that
\begin{equation}
\lim_{L\rightarrow\infty}\text{Pr}\left\{\ell_{2}<x\right\}
\sim\lim_{L\rightarrow\infty}\text{Pr}\left\{\ell_{1}<x\right\}.
\end{equation}
This follows from the fact that the expression $\text{Pr}\left\{\ell_{2}<x<\ell_{1}\right\}$ vanishes faster than $\text{Pr}\left\{\ell_{1}<x\right\}$ when $L\rightarrow\infty$. One can use the same approach in order to show that 
\begin{equation}
\lim_{L\rightarrow\infty}\text{Pr}\left\{\ell_{k+1}<x\right\}
\sim\lim_{L\rightarrow\infty}\text{Pr}\left\{\ell_{k}<x\right\}\ \ \ \forall k\ll L.
\end{equation} 

The dependency of the probability $\text{Pr}\left\{\ell_{2}<x<\ell_{1}\right\}$ on the length of the system is reflected in Figure 4 where the quantity
\begin{equation}
\mathcal{N}=\int \mathrm{d}\rho \, \text{Pr}\left\{\ell_{2}<1<\ell_{1}\right\}  
\end{equation}
is shown as function of the length of the system. We choose $\mathcal{N}$ in order to show that the likelihood of a transition $\{ \ell_{1}(t)>1\}\leftrightarrow\left\{\ell_{1}(t+1)<1\right\}$, which strongly depends on $\text{Pr}\left\{\ell_{2}<x<\ell_{1}\right\}$, is vanishing for all densities.

 \begin{figure}
	\input{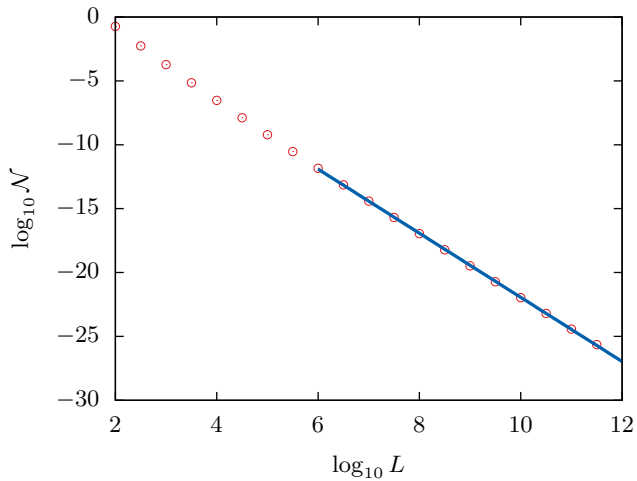}
	\caption{$\mathcal{N}=\int \mathrm{d}\rho \, \text{Pr}\left\{\ell_{2}<1<\ell_{1}\right\}$ vs the length of the system $L$. We find that for increasing lengths this weight tends to zero with an algebraic law. Numerically we find $\mathcal{N}\propto L^{-2.5}$ for $L\rightarrow\infty$. The blue line correspond to our numerical fit.}  
	\label{transition}
\end{figure}
\begin{figure}
	\input{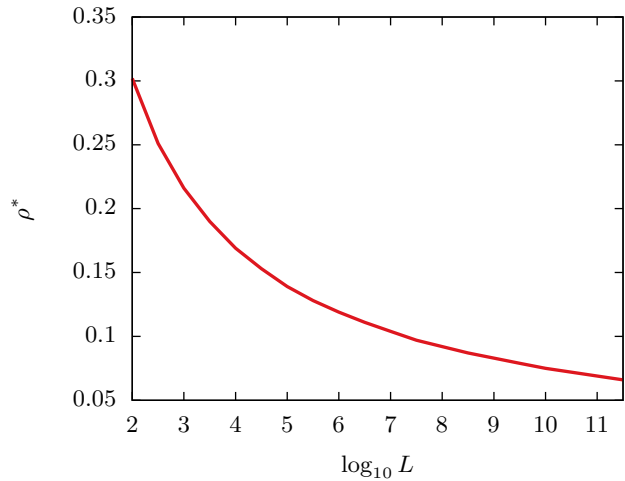}
	\caption{Density $\rho^{*}$ as function of the length of the system.}
		\label{fig5}
\end{figure}

On the other side if we consider the density $\rho^{*}$ for which the expression $\text{Pr}\left\{\ell_{2}<1<\ell_{1}\right\}$ is maximal,
\begin{equation}
\frac{d}{d \rho} \text{Pr} \left\{\ell_{2}<1<\ell_{1}\right\}|_{\rho^{*}}=0,
\end{equation}
we can see in Figure \ref{fig5} that the behavior of the function $\rho^{*}(L)$ is described by a monotone decreasing function for increasing lengths.

In detail we show that for sufficiently high densities $(\rho>\left(\ln L\right)^{-1})$ in the thermodynamic limit we arrive at a state where a macroscopic number of sites have a mass above $1$. If we assume that these states have a survival probability (meaning the probability of remaining above $1$) that is higher than zero, then the existence of a macroscopic number of such states in combination with the increasing survival probabilities for increasing masses when truncation dynamics apply is a sufficient condition for the creation of a condensate in the system.   

Similarly for $\rho\ll \left(\ln L\right)^{-1}$ we almost surely can observe states with $\ell_{1}(t)<1$ and a vanishing transition probability $\left\{\ell_{1}(t)<1\right\} \rightarrow \left\{\ell_{1}(t+1)>1\right\}$ for $L\rightarrow \infty$ and $\forall t\in\mathbb{N}$. By taking into  account these two facts the broken ergodicity property for truncated models in the thermodynamic limit becomes evident. 

\section{Conclusion}

We introduced and studied three different truncated random average processes. We started with the analysis of finite size systems which could be approached only by numerical methods. A convenient choice for the characteristic variables of this system are the largest single-site mass and the flow in the system. The introduction of the parameter $\gamma$ allowed us to control effectively the dynamics of the system and hereby the stability of the condensates that appear in the evolution of the system. The impact of this parameter on the condensation transition was studied by determining numerically the phase diagram in the $\rho-\gamma$ plane. By comparing the different diagrams we could also clarify the relevance of the different processes like prohibition of transport ($\delta$ - term) or reduction of the fraction ($\Theta$ - term) for the nature of the condensates.  

We were able to derive analytical results only in the limit $L\rightarrow\infty$ by studying the order statistics of the free ARAP. This unconvential approach proves extremely useful when dealing with problems of condensation transitions arising out of single site deviations. Specifically we could see that in the thermodynamic limit the system can be in either of two states: no site has a mass above the cutoff or a macroscopic number of sites has a mass above the cutoff. This property in combination with the stability of the condensates which could be observed by the numerical approach of section \textbf{III} confirm the broken ergodicity property described in \cite{Zielen023}.
      
Interesting generalizations of the presented model may arise. As example one could consider the transport properties of a similar model with open boundary conditions. Another  question that arose during this work and has not been answered yet is the behavior of systems with $\gamma=-\infty$. In this case an absorbing stationary state exists in the system and the relation to similar non-equilibrium processes is at hand. Surprisingly, as we will show in a future paper, it is possible to determine the single site mass distribution for this absorbing state analytically. Lastly it would be extremely interesting to consider the question with regard to the nature of the phase transition occurring in the crossover from the \textbf{HLF2} to the \textbf{C} phase.

\end{document}